# Power Quality Event Recognition and Classification Using an Online Sequential Extreme Learning Machine Network based on Wavelets

Rahul Kumar Dubey

Robert Bosch Engineering and Business Solutions Ltd., Bengaluru, India
*Tel- +918527457072
*E-mail: rahul.dubey2011@ieee.org

*Abstract*— Reduced system dependability and higher maintenance costs may be the consequence of poor electric power quality, which can disturb normal equipment performance, speed up aging, and even cause outright failures. This study implements and tests a prototype of an Online Sequential Extreme Learning Machine (OS-ELM) classifier based on wavelets for detecting power quality problems under transient conditions. In order to create the classifier, the OSELM-network model and the discrete wavelet transform (DWT) method are combined. First, discrete wavelet transform (DWT) multi-resolution analysis (MRA) was used to extract characteristics of the distorted signal at various resolutions. The OSELM then sorts the retrieved data by transient duration and energy features to determine the kind of disturbance. The suggested approach requires less memory space and processing time since it can minimize a large quantity of the distorted signal's characteristics without changing the signal's original quality. Several types of transient events were used to demonstrate the classifier's ability to detect and categorize various types of power disturbances, including sags, swells, momentary interruptions, oscillatory transients, harmonics, notches, spikes, flickers, sag swell, sag mi, sag harm, swell trans, sag spike, and swell spike.

1. Introduction

"The notion of powering and grounding electronic equipment's in a way that is suited for the functioning of that equipment and compatible with the premise wiring system and other connected equipment," says the IEEE 1100 Authoritative Dictionary. Electric power is becoming more polluted due to the proliferation of non-linear electric power systems and new power electronic gadgets. Misfunction, short lifespan of equipment, additional heating, and failure of end-use equipment are all direct results of power quality (PQ) events such voltage sag, swell, notches, flicker, glitches, and transients. Damage to our equipment isn't the only potential cost of low power quality (PQ) in commercial and industrial settings, such as a factory's quality control department; such a failure would prevent the load from being continued, disrupting the production process and potentially costing customers in a variety of ways [1–7]. The switching of an aluminum smelter, generation step load application, at the refinery due to utility fault, flicker from the operation of elevators, due to arc furnaces and arc welders, voltage swell due to step load rejection, and notches which are produced due to rectifiers, inverters, electric drives w. are all examples of Power Quality Disturbances(PQD). This combination of factors may lead to more than one PQ event occurring in certain cases. To automatically identify disturbances, take actions to reduce them, provide clean electricity to customers, and increase PQ for disturbance network protection, it is crucial that these PQ be monitored constantly.

Time and frequency resolution is extracted from PQD [2] using a variety of digital signal processing methods, including as the Fourier transform (FT), the Short time Fourier transform (STFT), the Wavelet transform (WT), and the Hilbert transform (HT). One of FT's drawbacks is that although it provides enough detail in the frequency domain, it provides little in the time domain. Due to the FT's resolution dependence on the window size, it is not useful for monitoring PQ signals, which are often non-stationary. Time information is not provided concurrently by Fourier analysis since it is lost in the transformation to the frequency domain. The non-stationary signals are decomposed into a time-frequency domain via the Short Time Fourier Transform (STFT), which employs a sliding window. Here, the frequency resolution in time is totally up to the window sizes. The signal's frequency and timing details cannot be determined at the same time, per Heisenberg's Uncertainty Principle (HUP). Since the window size is constant in STFT, the resolution is also constant. In order to deconstruct the signal at each scale and extract the feature to be utilized as inputs to neural network for classification, Santoso et.al [3] presented the usage of square integrable function (SWTC) and group theory representation. The aforementioned feature is used for classification, and it was first described by Agrisani et al. [4], who developed the iDiscrete Time Wavelet Transform (DTWT) to acquire the disturbance amplitude that contains a large amount of noise component from the signal through the Continuous Wavelet Transform (CWT). Standard deviation and RMS value are used as characteristics for classification in the wavelet multi-resolution signal decomposition approach introduced by Gaouda et al. [7]. WT and Probabilistic Neural Network (PNN) have been combined by Gaing [5] for PQD classification. Through the use of Parseval's theorem, he calculated the average power and energy distribution values of the disturbance signal in order to extract its defining characteristics. He has used these methods to try to minimize the size of the feature matrix by getting rid of the superfluous ivalues without losing any temporal information. The ST and PNN theory was utilized by Mishra et al. [13] to extract features from the phase and contour of the signal's S-matrix, which were then fed into a PNN-trained network that efficiently detected and localized PQD. Furthermore, they have compared PNN's performance to that of the LVQ and FFML neural networks. For correct categorization of the disturbance's type, WT [6] relies on its most notable feature: the enhancement (i.e., scaling and transformation) of the window's width for each spectral component.

The purpose of this paper is to categorize 16 distinct PQD. We have looked at both single and multiple Power Quality events, totaling eight categories, in this work. Events include sag, swell, temporary interruption, oscillatory transients, harmonics (3rd, 5th, 7th, 9th), notch, spike, flicker, sag swell, sag mi, sag harm, swell trans, sag spike, swell spike, and many more. Our primary goal in writing this research was to provide Online Sequential Extreme Learning Machine Learning for the classification of PQD (OS-ELM). In order to extract statistical and non-statistical characteristics at different frequency resolved sub-bands, the original signal is first treated using DWT, where the PQD are decomposed upto 11th level using db4 wavelet, and then rebuilt. Energy, skewness, mean, kurtosis, shanon entropy, and standard deviation are some of the retrieved characteristics. To do this, we built a feature set with 66 features (11 tiers of decomposition times 6 features per event). The larger size of the features acquired in this way improves classification accuracy but makes further calculation difficult and time-consuming. As a result, the OSELM was fed data that had been prepared for both training and testing as input by randomly sorting the feature set's dimensions. In this study, we examine the accuracy and computation time of several classification methods, including PNN and multiclass SVM (using 1 vs. 1), to determine which one is most suited for evaluating the PQD automatically and reliably detecting and localizing disturbances. The steps of a power quality analysis are shown in a block diagram in Figure 1.

Section 2 gives a high-level summary of WT technology, while Section 3 shows an example of feature extraction from simulated distorted PQ events. The OSELM methodology's underlying body of knowledge is presented in Section 4. The results of the PQD classification using PNN, SVM, and OSELM with various activation functions are discussed in Section 5. A summary is provided in Section 6.

2. **Wavelet theory**

The Wavelet Transform (WT) is a tool for carving up functions, operators, or data into components of different frequency, allowing one to study each component separately. Transients of high frequency present along with non-stationary and non-periodic signals can be efficiently analyzed using Wavelet Transform. It is also efficient in the analysis of harmonic signals. The basic idea of the wavelet transform is to represent any arbitrary function

$f(t)$ as a superposition of a set of such wavelets or basis functions at different positions and scales. These basis functions or baby wavelets are obtained from a single prototype wavelet called the mother wavelet, by dilations or contractions (scaling) and translations (shifts).The WT is computed separately for different segments of the time-domain signal at different frequencies. The wavelet transform is designed to give good time resolution but poor frequency resolution at high frequencies where as it gives good frequency resolution but poor time resolution at low frequencies. It is good for signals having high frequency components for short durations and low frequency components for long duration. There are two main functions that are used WT: wavelet function $\psi(t)$ and the scaling function $\phi(t)$ which perform the Multiresolution Analysis (MRA) decomposition as well as reconstruction of the signal[3]. The signal is decomposed into high frequency components (i.e. detailed version is generated from the wavelet function $\psi$) and the low frequency component (i.e. approximated version generated by the scaling function $\phi$) and retains the main features of original signal. The square integrable function and the group theory representation results in the decomposition into scales. WT is of basically two types Continuous Wavelet Transform (CWT) and Discrete Wavelet Transform (DWT).Continuous Wavelet coefficients are produced for different scaling and translation factor; thereby mapping a continuous function into a function of two continuous variable. DWT decomposes a discretized signal into different resolution levels which maps a sequence of number into different sequence of numbers. DWT transforms a discrete time signal to a discrete wavelet representation.

a) **Multiresolution Analysis (MRA)**

Multiresolution Analysis uses two filters a HPF and a LPF to decompose a given signal into a detailed version and a smoother version after down sampling it by a factor of 2 and the results of the LPF is used for decomposition into subsequent levels .The decomposed signal can be reconstructed back to original signal without negotiating with any time information[5]. The flowchart in Fig.2 illustrates the DWT and MRA in brief.

$$y(t) = CA_0 + CD_0$$
$$= CD_0 + CA_1 + CD_1$$
$$= CD_0 + CD_1 + CD_2 + CA_2$$
$$= CD_0 + CD_1 + CD_2 + \ldots \ldots .CD_n + CA_n \ldots \ldots \ldots \ldots \ldots \ldots \ldots \ldots \ldots \ldots \ldots \ldots \ldots \ldots \ldots \ldots \ldots (1)$$

Where,

$y(t)$ = original signal undergoing wavelet transformation;

$CA_i$ = approximated/smoothed version of the signal;

$CD_i$ = detailed version of the signal;

i = 1, 2, 3…n (levels of decomposition)

### b) Mathematical Modeling of DWT

The two basic functions scaling function $\phi(t)$ and wavelet function $\psi(t)$ are defined in DWT as:

$$\phi(t) = \sqrt{2}\sum_{k} h_k \phi(2t - k) \quad \ldots\ldots\ldots\ldots\ldots (2)$$

$$\psi(t) = \sqrt{2}\sum_{k} g_k \phi(2t - k) \quad \ldots\ldots\ldots\ldots\ldots (3)$$

The discrete sequences $h_k$ and $g_k$ represent discrete high pass and low pass filters which satisfy the condition $g_k=(-1)^k h_{N-1-k}$ and are unique to the chosen mother wavelet. The scaling and the wavelet function of the class of the orthonormal basis function can be defined as :-

$$\phi_{j,k}(t) = 2^{\frac{j}{2}} \phi(2^j t - k) \quad \ldots\ldots\ldots\ldots\ldots\ldots\ldots (4)$$

$$\psi_{j,k}(t) = 2^{\frac{j}{2}} \psi(2^j t - k) \quad \ldots\ldots\ldots\ldots\ldots\ldots\ldots (5)$$

Where j,k ∈ Z. j is the scaling or the dilation factor in time scale and amplitude, and b is the shifting or translation of the function in time using the above two equations (4),(5), the function y(t) can be easily expanded as:

$$y(t) = \sum_{j \in Z} c(l)\phi_l(t) + \sum_{j=0}^{J-1}\sum_{k=0}^{\infty} d(j,k)\psi_{j,k}(t) \quad \ldots\ldots\ldots\ldots\ldots\ldots (6)$$

y(t)= original signal, c(l) is the approximated coefficient and d(j,k) is the detailed coefficient which are calculated as inner product [8].

### c) Choice of wavelet

The choice of wavelet is very significant in analyzing, in the detection and localization of the Power Quality Disturbances. Each mother wavelet has its feasibility depending upon the application requirements. Daubechies wavelet is one of the most suitable basis function for analyzing power system's short as well as fast transients disturbance[3]-[6].

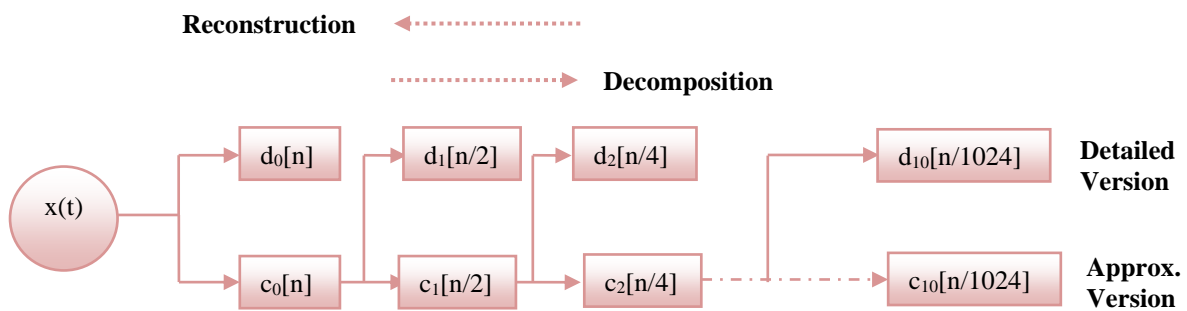

### Fig. 1. MRA using Wavelet Transform

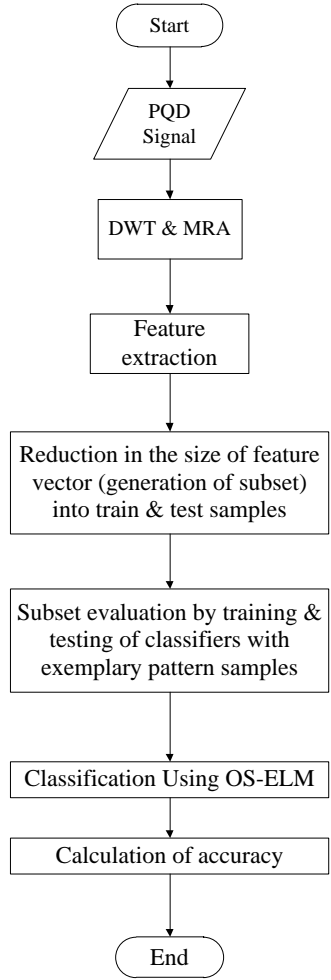

**Fig.2: Flowchart of the algorithm of PQ event classification**

**Table 1: Mathematical Model of PQ events**

| Events | Equation | Parameter |
|---|---|---|
| Normal signal | $y(t)=B\sin(\omega_0 t)$ | $\omega_0 = 2\pi f$; f=50Hz; B=amplitude |
| Sag(S1) | $y(t)= B(1-\alpha(u(t-t_a)-u(t-t_b)))\sin(\omega_0 t)$ | $0.16 < \alpha < 0.95$; $t_a < t_b$; $2T < t_b-t_a < 8T$; u(t)=1 t>0 =0 t<0 |
| Swell(S2) | $y(t)= B(1+\alpha(u(t-t_a)-u(t-t_b)))\sin(\omega_0 t)$ | $.05 < \alpha < .8$; $2T < t_b-t_a < 8T$ |
| Momentary interruption(S3) | $y(t)= B(1-\alpha(u(t-t_a)-u(t-t_b)))\sin(\omega_0 t)$ | $0.87 < \alpha < 1$; $2T < t_b-t_a < 8T$ |
| Oscillatory Transients(S4) | $y(t)=B \sin(\omega_0 t) + \alpha_0 \exp(-(t-t_a/\tau_{osc}))(u(t-t_a) - u(t-t_b))\sin(2\pi f_n t)$ | $10 < f_n < 100$Hz; $25 < \tau_{osc} < 100$; $\alpha_0=5$; $0.15T < t_b-t_a < 10T$ |
| Harmonics(S5) | $y(t)= B(\alpha_1\sin(\omega_0 t)+ \alpha_3\sin(3\omega_0 t)+(\alpha_5)\sin(5\omega_0 t)+ (\alpha_7)\sin(7\omega_0 t)+ \alpha_9\sin(9\omega_0 t))$ | $0 < \alpha_3\ \alpha_5\ \alpha_7\ \alpha_9 < 0.3$; $\alpha_1=1$; $\Sigma(\alpha_i)=1$ |
| Notch(S6) | $y(t)=(B\sin(\omega_0 t)+\text{sign}(B\sin(\omega_0 t)))*$ $[(\Sigma z*(u(t-(t_a+0.002m))-u(t-(t_b+0.002m))))]$ | $1<m<10; 0.1<z<0.4;$ $0<t_a;t_b<0.05T;$ $0.01T<t_b-t_a<0.05T$ |
| Spike(S7) | $y(t)=(B\sin(\omega_0 t)-\text{sign}(B\sin(\omega_0 t)))*$ $[(\Sigma z*(u(t-(t_a+0.002m))-u(t-(t_b+0.002m))))]$ | $1<m<10; 0.1<z<0.4;$ $0<t_a;t_b<0.05T;$ $0.01T<t_b-t_a<0.05T$ |
| Flicker(S8) | $y(t)=[1+\alpha\sin(2\pi\beta t)]B\sin(\omega_0 t)$ | $2<\beta<20$ Hz; $0.1<\alpha<0.2$ |
| Sag+Swell(S9) | $y(t)=B(1-\alpha_1(u(t-t_a)-u(t-t_b)))*$ $(1+\alpha_2(u(t-t_c)-u(t-t_d)))\sin(\omega_0 t)$ | $0.16<\alpha_1<0.95; .05<\alpha_2<.8;$ $2T<t_b-t_a<8T; 4T<t_d-t_c<9T$ |
| Sag+MI(S10) | $y(t)=B(1-\alpha_1(u(t-t_a)-u(t-t_b)))*$ $(1-\alpha_3(u(t-t_c)-u(t-t_d)))\sin(\omega_0 t)$ | $0.16<\alpha_1<0.95; 0.87<\alpha_3<1;$ $2T<t_b-t_a<8T; 4T<t_d-t_c<9T$ |
| Swell+MI(S11) | $y(t)=B(1-\alpha_2(u(t-t_a)-u(t-t_b)))*$ $(1+\alpha_3(u(t-t_c)-u(t-t_d)))\sin(\omega_0 t)$ | $.05<\alpha_2<.8; 0.87<\alpha_3<1;$ $2T<t_b-t_a<8T; 4T<t_d-t_c<9T$ |
| Sag+Transient(S12) | $y(t)=B(1-\alpha(u(t-t_a)-u(t-t_b)))\sin(\omega_0 t)+\alpha_0\exp(-(t-t_a/\tau_{osc}))(u(t-t_a)-u(t-t_b))\sin(2\pi f_n t)$ | $0.16<\alpha<0.95; 10<f_n<100$Hz; $25<\tau_{osc}<100; \alpha_0=5; 2T<t_b-t_a<8T$ |
| Swell+Tranients(S13) | $y(t)=B(1+\alpha(u(t-t_a)-u(t-t_b)))\sin(\omega_0 t)+\alpha_0\exp(-(t-t_a/\tau_{osc}))(u(t-t_a)-u(t-t_b))\sin(2\pi f_n t)$ | $.05<\alpha<.8; 10<f_n<100$Hz; $25<\tau_{osc}<100; \alpha_0=5; 2T<t_b-t_a<8$ |
| Sag+harmonics(S14) | $y(t)=B(1-\alpha(u(t-t_a)-u(t-t_b)))\sin(\omega_0 t)+B(\alpha_1\sin(\omega_0 t)+\alpha_3\sin(3\omega_0 t)+(\alpha_5)\sin(5\omega_0 t)+(\alpha_7)\sin(7\omega_0 t)+\alpha_9\sin(9\omega_0 t))$ | $0.16<\alpha<0.95; 0<\alpha_3\alpha_5\alpha_7\alpha_9<0.3;$ $\alpha_1=1;\Sigma(\alpha_i)=1; 2T<t_b-t_a<8T$ |
| Swell+harmonics(S15) | $y(t)=B(1+\alpha(u(t-t_a)-u(t-t_b)))\sin(\omega_0 t)+B(\alpha_1\sin(\omega_0 t)+\alpha_3\sin(3\omega_0 t)+(\alpha_5)\sin(5\omega_0 t)+(\alpha_7)\sin(7\omega_0 t)+\alpha_9\sin(9\omega_0 t))$ | $.05<\alpha<.8; 0<\alpha_3\alpha_5\alpha_7\alpha_9<0.3;$ $\alpha_1=1; \Sigma(\alpha_i)=1$ $2T<t_b-t_a<8T$ |
| Harmonics+transient(S16) | $B(\alpha_1\sin(\omega_0 t)+\alpha_3\sin(3\omega_0 t)+(\alpha_5)\sin(5\omega_0 t)+(\alpha_7)\sin(7\omega_0 t)+\alpha_9\sin(9\omega_0 t))+\alpha_0\exp(-(t-t_a/\tau_{osc}))(u(t-t_a)-u(t-t_b))\sin(2\pi f_n t)$ | $0<\alpha_3\alpha_5\alpha_7\alpha_9<0.3; \alpha_1=1;$ $\Sigma(\alpha_i)=1; 10<f_n<100$Hz; $25<\tau_{osc}<100; \alpha_0=5; 0.15T<t_b-t_a<10T$ |

In this paper, db4 has been used as it is the most localized rather compactly supported in time. db4 is orthogonal with highest number of vanishing moment for a given width. Besides being is shift variant in nature. These wavelets best match to signals with sag, swell or harmonic disturbances and hence are considered for feature extraction. Fig.2 shows a flowchart describing the entire process of PQ classification. The wavelet transformation using DWT and MRA forms the preprocessing part, the feature extraction and the training and testing of the samples presents the processing part and the classification using OSELM and the determination of the accuracy is the post processing of the given signal which results in accurate assessment of PQD so as to detect and localize it automatically.

3. **Feature Extraction**

Synthetic PQ distorted signals were generated using MATLAB as shown in Table 1 which describes the mathematical model of the generated signals. The signal was generated with a sampling frequency of (256*50) 12.8 kHz. Ten cycles of signal was considered for analysis at 2560 points. The signal was decomposed upto 11th level. Figure 3 to Fig.7 shows the some of the PQD signals generated based on Table-2 and there corresponding wavelet decomposition db3. The detailed coefficient $CD_{ij}$ at each decomposition level is used to extract the statistical and non-statistical features like energy($EDR_i$), skewness($SKW_i$), mean($\mu_i$), shanon entropy($ENTP_i$), kurtosis ($KRT_i$) and standard deviation ($\sigma_i$). The equations used in Table 2 is used to calculate the above mentioned features of the distorted signal at each decomposition level. The feature vector of length 6x11 i.e. 66 was constructed as follows:

Table 2: Formula for Feature Extraction of Detailed Coefficients [8]

| Sl.no. | Features | Features Formulation (i=1, 2, 3….11) |
|---|---|---|
| 1 | Energy | $EDR_i = \sum_{j=1}^{N} 1 \cdot \left|CD_{ij}\right|^2$ |
| 2 | Standard Deviation | $\sigma_i = \left[\dfrac{1}{(N-1)} \sum_{j=1}^{N} 1 \cdot \left|CD_{ij} - \mu_i\right|^2\right]^{1/2}$ |
| 3 | Mean | $\mu_i = \dfrac{1}{N} \sum_{j=1}^{N} 1 \cdot \left|CD_{ij}\right|$ |
| 4 | Kurtosis | $KRT_i = \left[E \cdot (CD_{ij} - \mu_i)^4\right] / \sigma_i^4$ |
| 5 | Skewness | $SKW_i = \left[E \cdot (CD_{ij} - \mu_i)^3\right] / \sigma_i^3$ |
| 6 | Shanon Entropy | $ENTP_i = -\sum_{j=1}^{N} 1 \cdot CD_{ij}^2 \log(CD^2_{ij})$ |

In Table 2, N is the number of coefficients in each decomposed data. The term $E(CD_{ij} - \mu_i)^k$ in skewness and kurtosis calculation is the expected value of the quantity, also called as $k^{th}$ moment about the mean. Thus in the present case the decomposition of the signal upto $11^{th}$ level yields 6 different nodes corresponding to different frequency sub-bands. Selecting the above mentioned features for each node we will obtain the feature vector of length 6.

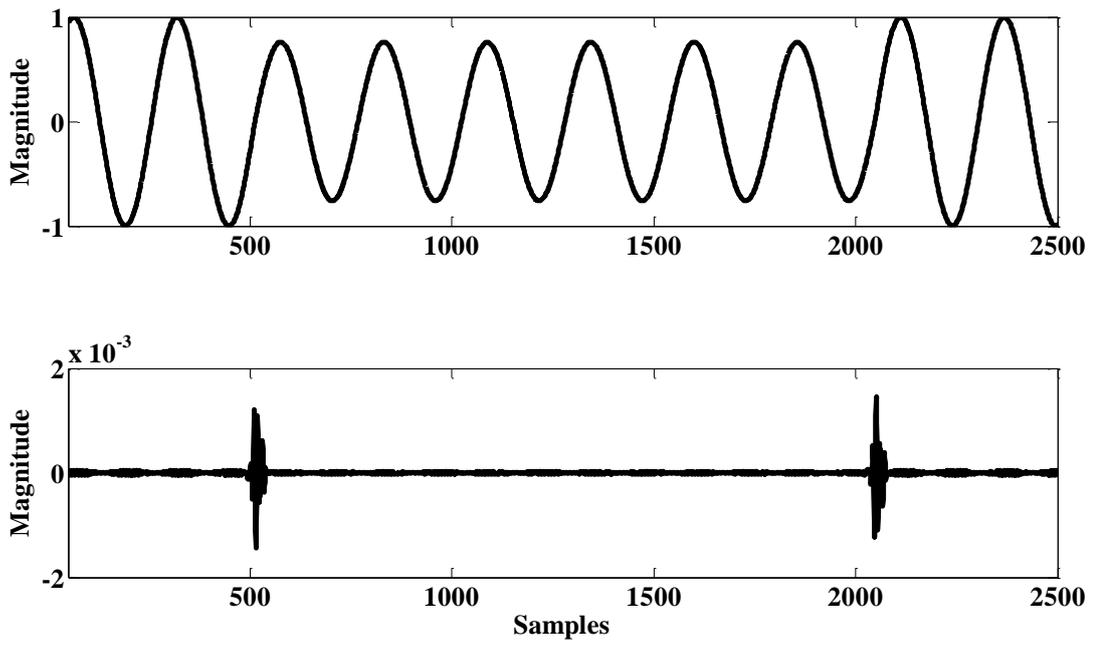

**Fig 3: (a) voltage sag and (b) 3rd level WT decomposition of voltage sag**

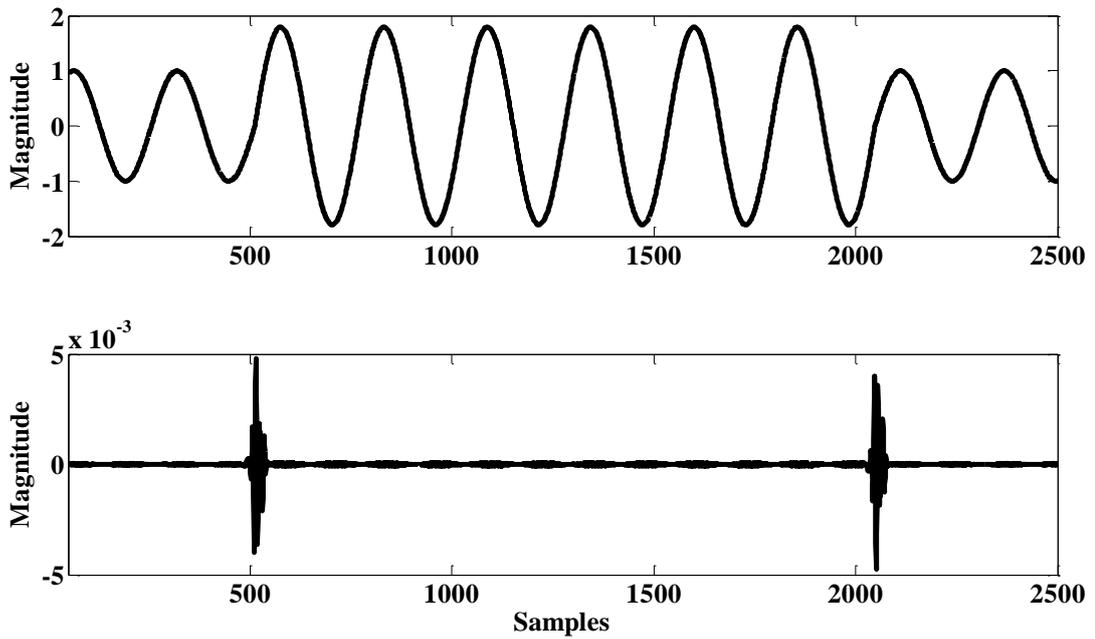

**Fig 4: (a) voltage swell and (b) 3rd level WT decomposition of voltage swell**

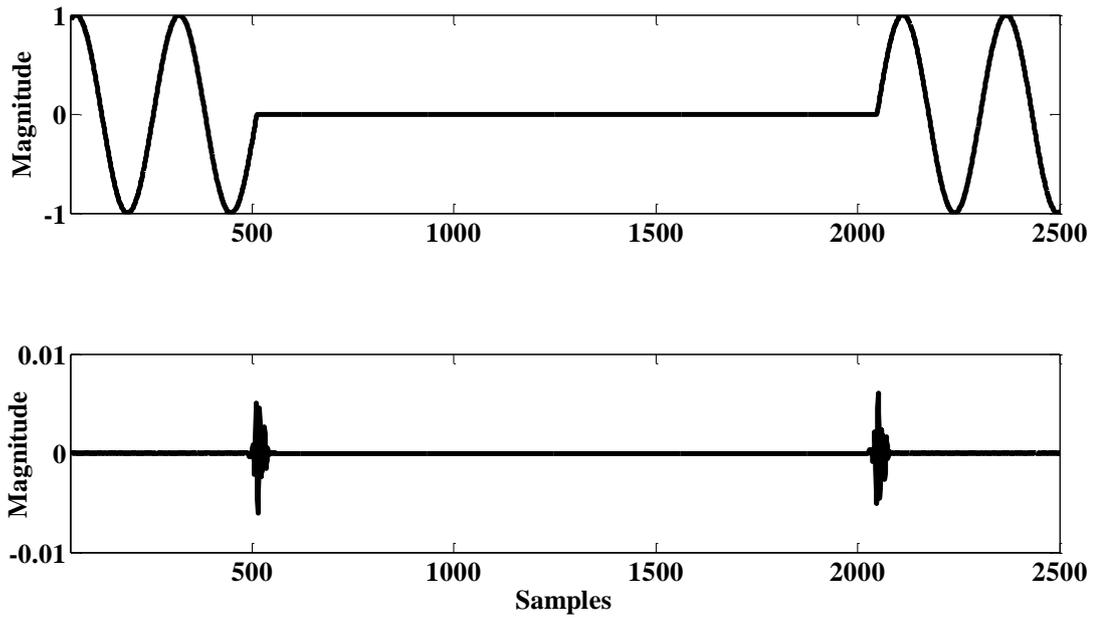

**Fig 5: (a) momentary interruption and (b) 3rd level WT decomposition of momentary interruption**

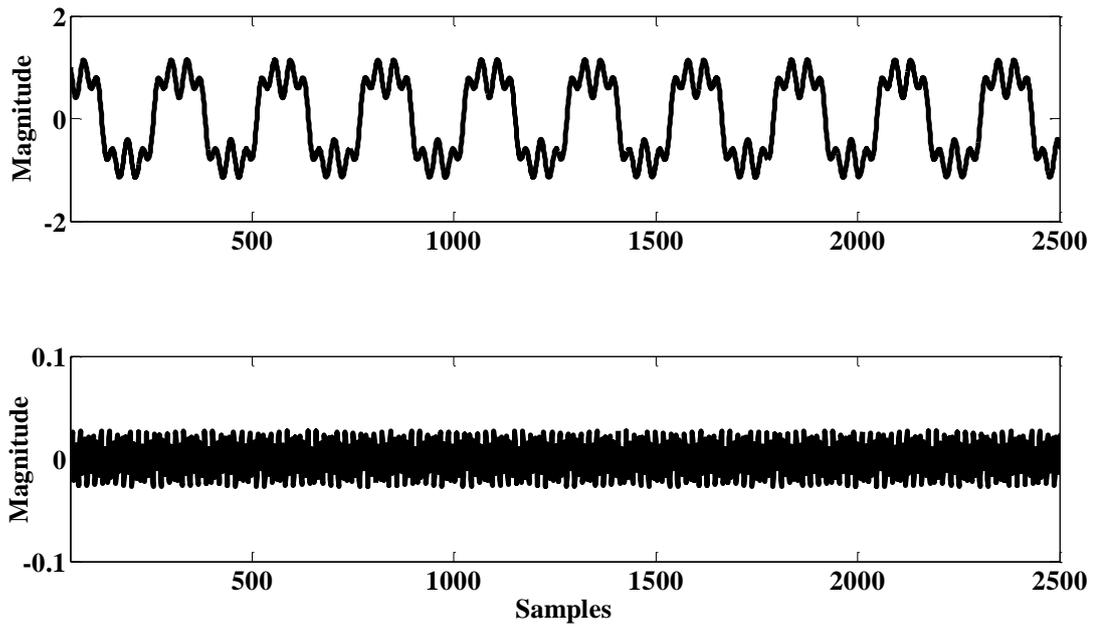

**Fig 6: (a) harmonics (3rd, 5th, 7th, and 9th) and (b) 3rd level WT decomposition of harmonics**

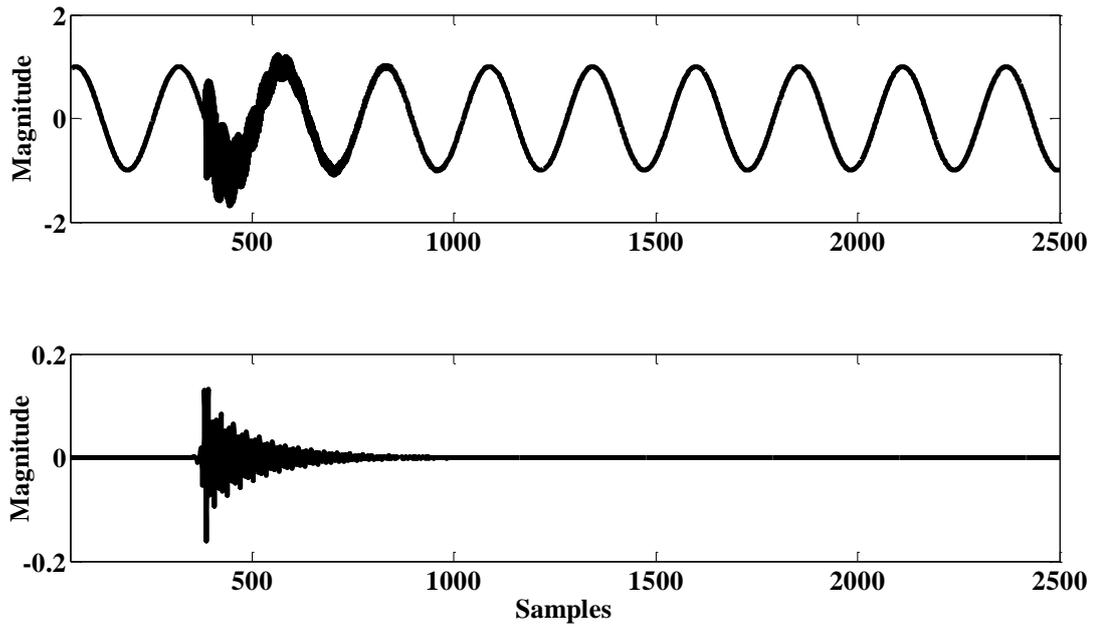

**Fig 7: (a) oscillatory transient and (b) 3rd level decomposition**

### 4. Proposed Classifier:-OSELM

The algorithm is referred to as online sequential extreme learning machine (OS-ELM) and can learn data one-by-one or chunk-by-chunk (a block of data) with fixed or varying chunk size. The activation functions for additive nodes in OS ELM can be any bounded non-constant piecewise continuous functions and the activation functions for RBF nodes can be any integrable piecewise continuous functions. In OS-ELM, the parameters of hidden nodes (the input weights and biases of additive nodes or the centers and impact factors of RBF nodes) are randomly selected and the output weights are analytically determined based on the sequentially arriving data. The algorithm uses the ideas of ELM of Huang *et al.* developed for batch learning which has been shown to be extremely fast with generalization performance better than other batch training methods. Apart from selecting the number of hidden nodes, no other control parameters have to be manually chosen. Detailed performance comparison of OS-ELM is done with other popular sequential learning algorithms on benchmark problems drawn from the regression, classification and time series prediction areas. The results show that the OS-ELM is faster than the other algorithms and produces better generalization performance**.** It is a feed forward incremental learning network without having to add hidden neurons [12]. In the initial stage, the OSELM requires a fixed number of hidden neurons. The input weights are randomly generated too. Both input weights and the number of hidden neurons remain unchanged after that. During the training cycle, an initial training phase is required, and a set of data samples is to be presented (as a group) to the network, before other data samples are presented on a one-by-one basis. The number of data samples for the initial training phase is equal to or larger than the number of hidden neurons. The output weights and bias are updated based on the Sherman-Morrison formula. The OSELM is an extended version of the batch-mode training procedure of the ELM .Figure 8 given below shows the architecture of OSELM.OSELM is considered as feed forward or a RBF (depends to type of activation function), with advanced learning algorithm. Consider a set of $N$ training samples (with a input vector and respectively target output vector), $(\mathbf{x}_j, \mathbf{t}_j) \in \mathbf{R}^n \times \mathbf{R}^m$, is used to training OSELM that with $L$ number of hidden nodes. In a perfect case, the output of this OSELM respectively to $\mathbf{x}_j$ should be;

$$f(x_j) = \sum_{i=1}^{L} \beta_i G(a_i, b_i, x_j) = t_j \text{ For } j = 1 \dots N \qquad (7)$$

Where $\mathbf{a}_i$ and $\mathbf{b}_i$ are the input weights and bias (learning parameters) of the hidden nodes, $b_i$ is the output weights, and $G(\mathbf{a}_i, b_i, \mathbf{x}_j)$ is the output of the $i^{th}$ hidden neuron respectively to the input vector $\mathbf{x}_j$.

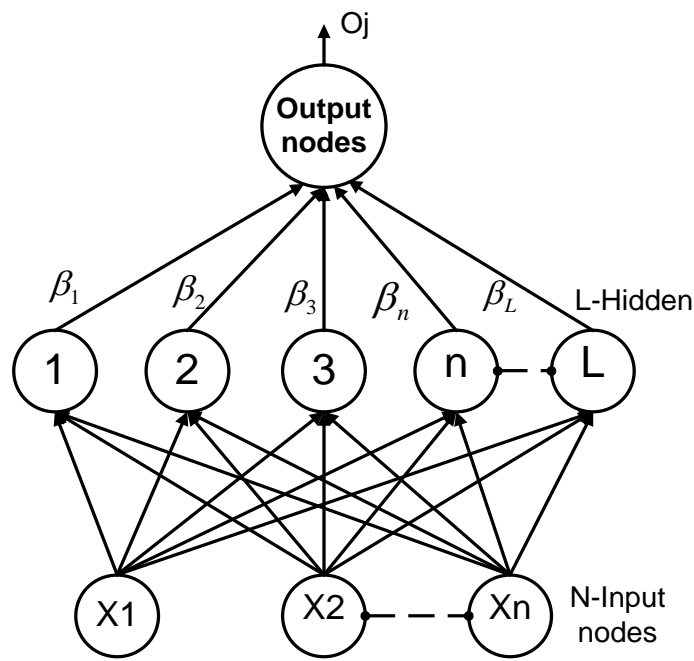

**Fig.8: Architecture of OSELM**

Equation (8) and (9) show the definition of the $G(\mathbf{a}_i, b_i, \mathbf{x}_j)$ for additive hidden neuron and RBF neuron, respectively. There are two phases in the training of OSLEM, i.e., the initialization phase and sequential learning phase. In the initialization phase, a small chunk of training data (denoted as $N_0$) to train the OSELM with $L$ (where $N_0 >= L$) with hidden neurons[13]. The procedures of the initialization phase are as follows.

**Step 1: Initialization phase:**

Initialize the learning using a small chunk of initial training data $\alpha_0 = \{(x_i, t_i)\}_{i=1}^{N_0}$ from the given training set

$$\alpha = \{(x_i, t_i) | x_i \in R^n, t_i \in R^m, i = 1, 2, \ldots N\}, N_0 \geq \tilde{N}.$$

   a) Assign random input weights $a_i$ and bias $b_i$ (for additive hidden nodes) or center $a_i$ and impact factor $b_i$ (for RBF hidden nodes), $i = 1, 2, \ldots N$.

   b) Calculate the initial hidden layer output matrix $H_0$:

$$H_0 = \begin{bmatrix} G(a_1, b_1, x_1) & \cdots & G(a_{\tilde{N}}, b_{\tilde{N}}, x_1) \\ \vdots & \cdots & \vdots \\ G(a_1, b_1, x_{N_0}) & \cdots & G(a_{\tilde{N}}, b_{\tilde{N}}, x_{N_0}) \end{bmatrix}_{N_0 \times \tilde{N}} \quad (8)$$

   c) Estimate the initial output weight: $\beta^{(0)} = P_0 H_0^T T_0,$ \hfill (9)

   Where, $P_0 = (H_0^T H_0)^{-1}$ and $T_0 = [t_1, t_2, \ldots, t_{N_0}]^T$.

   d) Set $k = 0$, where $k$ is the number of chunks that is trained currently.

**Step 2: Sequential learning phase:**

   (a) Present the $(k+1)^{th}$ chunk of new observations, $\alpha_{k+1} = \{(x_i, t_i)\}_{i=\left(\sum_{j=0}^{k} N_J\right)+1}^{\left(\sum_{j=0}^{k+1} N_J\right)}$

   Where, $N_{k+1}$ denotes the number of observations in $(k+1)^{th}$ chunk.

   (b) Calculate the partial hidden layer output matrix $H_{k+1}$ for the $(k+1)^{th}$ chunk of data $\alpha_{k+1}$

$$H_{k+1} = \begin{bmatrix} G\left(a_1, b_1, x_{i=\left(\sum_{j=0}^{k} N_J\right)+1}\right) & \cdots & G\left(a_{\tilde{N}}, b_{\tilde{N}}, x_{i=\left(\sum_{j=0}^{k} N_J\right)+1}\right) \\ \vdots & \cdots & \vdots \\ G\left(a_1, b_1, x_{i=\left(\sum_{j=0}^{k+1} N_J\right)+1}\right) & \cdots & G\left(a_{\tilde{N}}, b_{\tilde{N}}, x_{i=\left(\sum_{j=0}^{k+1} N_J\right)+1}\right) \end{bmatrix}_{N_{K+1} \times \tilde{N}}$$

   (c) Set $T_{k+1} = \left[t_{i=\left(\sum_{j=0}^{k} N_J\right)+1}, \ldots, t_{i=\left(\sum_{j=0}^{k+1} N_J\right)+1}\right]^T$.

   (d) Calculate the output weight $\beta^{(k+1)}$

$$P_{k+1} = P_k - P_k H_{k+1}^T \left(I + H_{k+1} P_k H_{k+1}^T\right)^{-1} H_{k+1} P_k$$

$$\beta^{(k+1)} = \beta^{(k)} + P_k H_{k+1}^T \left(T_{k+1} - H_{k+1} \beta^{(k)}\right)$$

(e) Set $k = k+1$. Go to step 2(a).

As seen from the above OS-ELM algorithm, OS-ELM and ELM can achieve the same learning performance (training error and generalization accuracy) when $rank(H_0) = \tilde{N}$. The training data is received one-by-one instead of chunk-by-chunk $N_{K+1} \equiv 1$, step 2(d) has the following simple format (Sherman–Morrison formula[12-13]):-

$$P_{k+1} = P_k - \frac{P_k h_{k+1} h_{k+1}^T P_k}{1 + h_{k+1}^T P_k h_{k+1}}$$

$$\beta^{(k+1)} = \beta^{(k)} + P_{k+1} h_{k+1}^T \left(t_{k+1}^T - h_{k+1}^T \beta^{(k)}\right)$$

Where, $h_{k+1} = \left[G(a_1, b_1, x_{k+1}), \ldots, G(a_{\tilde{N}}, b_{\tilde{N}}, x_{k+1})\right]$

5. **Results and Discussion**

Sixteen classes (S1-S16) of different PQ disturbances (S1-S8 all are considered as a single disturbance and S9-S16 all considered as a mixed disturbance) are taken for classification and they are as follows:

S1→ Voltage Sag

S2→ Voltage Swell

S3→ Interruption

S4→ Oscillatory Transients

S5→ Harmonic Distortion

S6→ Notch

S7→Spike

S8→flicker

S9→ Sag with Swell

S10→ Sag with interruption

S11→ Swell with interruption

S12→ Sag with Transient

S13→ Swell with Transient

S14→ Sag with Harmonics

S15→Swell with Harmonics

S16→Harmonics with Transient

The power quality signals corresponding to these sixteen classes are generated in Matlab, using parametric models with different parameter values. The voltage sag / swell signals are generated by varying the sag / swell magnitude and duration as shown in Table 1. Harmonic signals are generated for different harmonic contents ($3^{rd}$, $5^{th}$, $7^{th}$, and $9^{th}$) and for different durations. Similarly the transient signals are generated for different duration and for different values of transient frequency. Besides these single disturbances, the following multiple disturbances are created for effective classification .A sag/swell is created with harmonic contents in the signal to generate sag/swell with harmonic. Other multiple disturbances as mentioned above are created by suitably choosing the parameters of the event. The sampling frequency of the signals are chosen as 12.8 kHz. 10 cycles of the data (2560 points) is considered for the analysis purpose. Wavelet transform of these data samples are then performed to decompose the signals up to $11^{th}$ level. The features from each of the decomposed levels (D1 to D11) constitute the feature vector as described in Table 2. Based on the extracted feature, the feature data sets for training and testing are constructed separately. The data set comprises of all six type of feature vectors for different types of disturbances. The classification accuracy of the data set is computed using SVM classifier for automatic classification of PQ events using each individual feature set corresponding to only one statistical / non-statistical measure as well as all 66 features. As a next step the OSELM is used to train the signals using these extracted features as input to the network in the initialization phase with some optimum number of hidden neuron and providing the data chunk-by-chunk to network. The number of training and testing sample used in each set of classes (i.e. 11, 13, 16 classes) have been shown in Table 3. Different kinds of activation functions were used like sigmoid, sinusoidal, rbf, hardlim to determine the training and testing accuracy and the estimate the computation time and the results were compared in Table 4. Amongst the various kinds of activation functions used sigmoid provides the best accuracy with minimum computation time followed by sinusoidal , rbf, and hardlim. The time taken by rbf is almost twice as that of sigmoid function whereas hardlim provides the least accuracy. The comparison has been made by taking into consideration first 11 classes (S1-S11), 13 classes (S1-S13), 16 classes (S1-S16).

Table 3: Specification of data set and the no. of training and testing samples

| Data set | Classes | Training samples | Testing samples |
|---|---|---|---|
| PQ events | 11 | 3254 | 815 |
| PQ events | 13 | 3510 | 879 |
| PQ events | 16 | 4353 | 1090 |

Table 4: Classification of PQD using OS-ELM

| Classes | Activation function | Timing (secs) | | Accuracy (%) | | Hidden neuron |
|---|---|---|---|---|---|---|
| | | Training | Testing | Training | Testing | |
| 11 | sigmoid | 2.6052 | 0.0624 | 99.85 | 99.63 | 500 |
| | rbf | 2.5584 | 0.1872 | 97.73 | 96.89 | |
| | sinusoidal | 3.1044 | 0.0624 | 99.91 | 99.51 | |
| | hardlim | 2.1372 | 0.0624 | 89.52 | 88.47 | |
| 13 | sigmoid | 3.0732 | 0.0624 | 99.77 | 99.43 | 500 |
| | rbf | 2.6208 | 0.436 | 97.01 | 95.90 | |
| | sinusoidal | 2.6832 | 0 | 99.72 | 99.09 | |
| | hardlim | 2.0436 | 0.0312 | 84.84 | 79.29 | |
| 16 | sigmoid | 5.4288 | 0.1092 | 99.93 | 99.72 | 700 |
| | rbf | 5.8500 | 0.1716 | 97.75 | 97.25 | |
| | sinusoidal | 6.2088 | 0.0624 | 99.93 | 99.17 | |
| | hardlim | 4.3524 | 0.0624 | 90.76 | 87.98 | |

To evaluate the performance of OSELM, its results are compared with the PNN network and a multiclass Support Vector Machines (SVM) using the one-vs.-one method of multiclass classification. A probabilistic neural network is predominantly a classifier which maps any input pattern to a number of classifications and can be forced into a more general function approximation. PNN is a supervised learning algorithm, but includes no weights in the hidden layer. PNN consists of an input layer, which represents input pattern or feature vector. The input layer is connected with hidden layer consisting of training layer [9].The actual vector serves as a weights as applied to input layer, finally output layer represents each of possible causes for which input layer can be classified. However, hidden layer does not connect to output layer. The support vector algorithm is based on statistical learning Support vector machines are designated for binary classification by creating a hyper plane to classify two different classes[10] but now multiclass classification is also possible using one-vs.-all approach, one-vs.-one approach, Dag SVM, Twin SVM which can effectively classify PQD[11] . At the outset, the performance analysis of PNN, SVM, and OSELM is carried out only for eleven classes i.e., (S1-S11) for the classification. Each neural network is trained with 80 % of the input data (or events) of each class and 20% of the data (or events) of each class are considered for testing. The classification results during testing are shown in Table 5. The overall accuracy of correct classification is the ratio of correctly classified events to that of the total number of events. The overall classification accuracy for PNN, SVM, and OSELM is 96.81%, 99.63%, and 99.63% respectively (Table 5). Hence, OSELM gives the best classification results for this case. With the same data, these networks are trained and subsequently tested for higher number of classes.

Table 5: Classification results of PNN, SVM, and OSELM for various number of classes

| Classes | Over all Accuracy (%) | | |
|---|---|---|---|
| | PNN | SVM | OSELM |
| 11 | 96.81 | 99.63 | 99.63 |
| 13 | 96.59 | 99.66 | 99.43 |
| 16 | 97.25 | 94.11 | 99.72 |

Table 6: Optimized values of hidden neurons for OSELM classification

| Sl. No. | No.of hidden neurons | Testing accuracy (%) | Training time(sec) |
|---|---|---|---|
| 1 | 50 | 85.78 | 0.1248 |
| 2 | 100 | 91.19 | 0.2496 |
| 3 | 150 | 94.68 | 0.4368 |
| 4 | 200 | 97.43 | 0.6240 |
| 5 | 250 | 97.98 | 0.8763 |
| 6 | 300 | 98.44 | 1.1856 |
| 7 | 350 | 98.53 | 1.6848 |
| 8 | 400 | 98.99 | 2.0124 |
| 9 | 450 | 99.08 | 2.2152 |
| 10 | 500 | 99.08 | 2.8080 |
| 11 | 550 | 99.27 | 3.6816 |
| 12 | 600 | 99.27 | 4.6020 |
| 13 | 700 | 99.36 | 6.2244 |
| 14 | 800 | 99.27 | 7.1136 |
| 15 | 900 | 99.27 | 9.9685 |
| 16 | 1000 | 99.27 | 13.0729 |

Table 7: Classification results of SVM for 16 classes

| | S1 | S2 | S3 | S4 | S5 | S6 | S7 | S8 | S9 | S10 | S11 | S12 | S13 | S14 | S15 | S16 |
|---|---|---|---|---|---|---|---|---|---|---|---|---|---|---|---|---|
| S1→ | **79** | 0 | 0 | 0 | 0 | 0 | 0 | 0 | 1 | 0 | 0 | 0 | 0 | 0 | 0 | 0 |
| S2→ | 0 | **76** | 0 | 0 | 0 | 0 | 0 | 0 | 0 | 0 | 0 | 0 | 0 | 0 | 0 | 0 |
| S3→ | 0 | 0 | **59** | 0 | 0 | 0 | 0 | 0 | 1 | 0 | 0 | 0 | 0 | 0 | 0 | 0 |
| S4→ | 0 | 0 | 0 | **60** | 0 | 0 | 0 | 0 | 0 | 0 | 0 | 0 | 0 | 0 | 0 | 0 |
| S5→ | 0 | 0 | 0 | 0 | **86** | 0 | 0 | 0 | 0 | 0 | 0 | 0 | 0 | 0 | 0 | 0 |
| S6→ | 0 | 0 | 0 | 0 | 0 | **126** | 0 | 0 | 0 | 0 | 0 | 0 | 0 | 0 | 0 | 0 |
| S7→ | 0 | 0 | 0 | 0 | 0 | 0 | **64** | 0 | 0 | 0 | 0 | 0 | 0 | 0 | 0 | 0 |
| S8→ | 0 | 0 | 0 | 0 | 0 | 0 | 0 | **20** | 0 | 0 | 0 | 0 | 0 | 0 | 0 | 0 |
| S9→ | 0 | 0 | 0 | 0 | 0 | 0 | 0 | 0 | **80** | 0 | 0 | 0 | 0 | 0 | 0 | 0 |
| S10→ | 0 | 0 | 0 | 0 | 0 | 0 | 0 | 0 | 0 | **80** | 0 | 0 | 0 | 0 | 0 | 0 |
| S11→ | 0 | 0 | 0 | 0 | 0 | 0 | 0 | 0 | 0 | 0 | **80** | 0 | 0 | 0 | 0 | 0 |
| S12→ | 0 | 0 | 0 | 0 | 0 | 0 | 0 | 0 | 0 | 0 | 0 | **32** | 0 | 0 | 0 | 0 |
| S13→ | 0 | 0 | 0 | 0 | 0 | 0 | 0 | 0 | 0 | 0 | 0 | 0 | **32** | 0 | 0 | 0 |
| S14→ | 0 | 0 | 0 | 0 | 0 | 0 | 0 | 0 | 0 | 0 | 0 | 0 | 0 | **86** | 0 | 0 |
| S15→ | 0 | 62 | 0 | 0 | 0 | 0 | 0 | 0 | 0 | 0 | 0 | 0 | 0 | 0 | **34** | 0 |
| S16→ | 0 | 0 | 0 | 0 | 0 | 0 | 0 | 0 | 0 | 0 | 0 | 0 | 0 | 0 | 0 | **32** |

**Overall Accuracy=94.11%**

Table 8: Classification results of PNN for 16 classes

| | S1 | S2 | S3 | S4 | S5 | S6 | S7 | S8 | S9 | S10 | S11 | S12 | S13 | S14 | S15 | S16 |
|---|---|---|---|---|---|---|---|---|---|---|---|---|---|---|---|---|
| S1→ | **75** | 0 | 0 | 0 | 0 | 0 | 0 | 0 | 2 | 3 | 0 | 0 | 0 | 0 | 0 | 0 |
| S2→ | 0 | **76** | 0 | 0 | 0 | 0 | 0 | 0 | 0 | 0 | 0 | 0 | 0 | 0 | 0 | 0 |
| S3→ | 0 | 0 | **61** | 0 | 0 | 0 | 0 | 0 | 0 | 0 | 0 | 0 | 0 | 0 | 0 | 0 |
| S4→ | 0 | 0 | 0 | **59** | 0 | 0 | 0 | 0 | 0 | 0 | 0 | 0 | 2 | 0 | 0 | 0 |
| S5→ | 0 | 0 | 0 | 0 | **87** | 0 | 0 | 0 | 0 | 0 | 0 | 0 | 0 | 0 | 0 | 0 |
| S6→ | 0 | 0 | 0 | 0 | 0 | **126** | 0 | 0 | 0 | 0 | 0 | 0 | 0 | 0 | 0 | 0 |
| S7→ | 0 | 0 | 0 | 0 | 0 | 0 | **64** | 4 | 0 | 0 | 0 | 0 | 0 | 0 | 0 | 0 |
| S8→ | 0 | 2 | 0 | 0 | 0 | 0 | 0 | **18** | 0 | 0 | 0 | 0 | 0 | 0 | 0 | 0 |
| S9→ | 5 | 8 | 0 | 0 | 0 | 0 | 0 | 0 | **63** | 0 | 0 | 0 | 0 | 0 | 0 | 0 |
| S10→ | 2 | 0 | 0 | 0 | 0 | 0 | 0 | 0 | 0 | **78** | 0 | 0 | 0 | 0 | 0 | 0 |
| S11→ | 0 | 0 | 0 | 0 | 0 | 0 | 0 | 0 | 0 | 0 | **80** | 0 | 0 | 0 | 0 | 0 |
| S12→ | 0 | 0 | 0 | 0 | 0 | 0 | 0 | 0 | 0 | 0 | 0 | **32** | 0 | 0 | 0 | 0 |
| S13→ | 0 | 0 | 0 | 2 | 0 | 0 | 0 | 0 | 0 | 0 | 0 | 0 | **30** | 0 | 0 | 0 |
| S14→ | 0 | 0 | 0 | 0 | 0 | 0 | 0 | 0 | 0 | 0 | 0 | 0 | 0 | **87** | 0 | 0 |
| S15→ | 0 | 0 | 0 | 0 | 0 | 0 | 0 | 0 | 0 | 0 | 0 | 0 | 0 | 0 | **62** | 0 |
| S16→ | 0 | 0 | 0 | 0 | 0 | 0 | 0 | 0 | 0 | 0 | 0 | 0 | 0 | 0 | 0 | **62** |

**Overall Accuracy=97.25%**

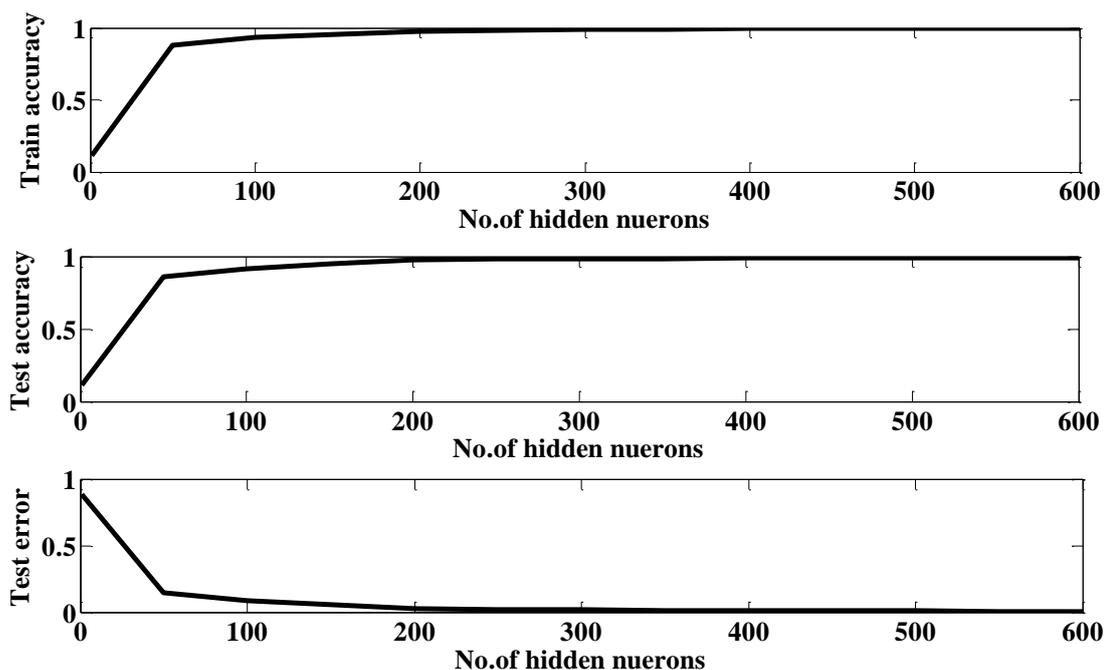

**Fig.9: (a) Plot of no.of hidden neurons vs. train_acc; (b) Plot of no.of hidden neurons vs. test_acc and (c) Plot of no.of hidden neurons vs. test_error**

The testing results are shown in Table 5. From Table 5, it is seen that as number of classes increases the overall classification accuracy of PNN and SVM degrades .Though for 13 classes the testing accuracy of SVM is slightly greater than OSELM, but the computation time of SVM is very high. On the other hand, classification performance of OSELM is excellent even for 16 number of classes. Here, OSELM is optimized for number of hidden neurons in all the cases and statistics is given in Table 6. The individual and overall classification for SVM and PNN with 16 number of classes are shown in Tables 7 and 8, respectively. The graphs in Fig.9 show the plot of number of hidden neurons vs. training accuracy (a), testing accuracy (b), testing error(c). The classification accuracy of the PQD with different classifiers and different transform is compared and presented in Table 9. The proposed classifier is able to classify 16 PQ disturbances with an accuracy of 99.72% which is higher than others. These obtained results show a satisfactory performance of this algorithm for classifications of various single-stage and multistage PQ disturbances.

**Table 9: Comparative assessment of the proposed method with other classifiers**

| References | Transform | Classifier | Type of data | No.of classes | Accuracy (%) |
|---|---|---|---|---|---|
| **[7]** | WT | MED | syn | 6 | 60.50 |
| **[5]** | WT | PNN | syn | 6 | 90.00 |
| **[16]** | ST | Fuzzy | syn | 7 | 98.00 |
| **[15]** | ST | PNN | syn | 10 | 94.7 |
| **[14]** | ST | PNN | syn | 11 | 97.4 |
| **[17]** | ST | Fuzzy | real | 13 | 92.7 |
| **Proposed** | WT | OSELM | syn | 16 | 99.72 |

## 6. Conclusion

In this paper an attempt has been made to improve the overall performance of a OSELM classifier using an optimal structure of the various activation functions. Sigmoid, rbf, sinusoidal algorithm has been used for that purpose and results corroborate the feature extraction mechanism showing high classification accuracy of PQ disturbance at reduced complexity. It is observed that the OSELM correctly classifies the PQ class with high accuracy. The OSELM is compared with both PNN and SVM and it is found that OSELM gives the best result. Therefore, the proposed method can be used as an effective power quality event classifier.

## 7. Acknowledgements

This research work is supported by the Prime Minister's Fellowship for Doctoral Research and being implemented jointly by Science & Engineering Research Board (SERB) and Confederation of Indian Industry (CII), with industry partner Robert Bosch.

## 8. References

[1] Power Quality, C. SANKARAN by CRC PRESS Boca Raton London New York Washington, D.C.

[2] RobiPolikar, The Engineer's Ultimate Guide to Wavelet analysis, The Wavelet Tutorial. March 1999.

[3] S. Santoso, E. J. Powers, W. M. Grady, and P. Hofmann, "Power quality assessment via wavelet transform analysis," *IEEE Trans.Power Del.*, vol. 11, no. 2, pp. 924–930, Apr. 1996.

[4] L. Angrisani, P. Daponte, M.D'Apuzzo, and A. Testa, "A measurement method based on the wavelet transform for power quality analysis," *IEEE Trans. Power Del.*, vol. 13, no. 4, pp. 990–998, Oct. 1998.

[5] Z. L. Gaing, "Wavelet-Based neural network for power disturbance recognition and classification," *IEEE Trans. Power Del.*, vol. 19, no. 4, pp. 1560–1568, Oct. 2004

[6] S. Santoso, E. J. Powers, W. M. Grady, and A. C. Parsons, "Power quality disturbance waveform recognition using wavelet-based neural classifier-part 1: Theoretical foundation," *IEEE Trans. Power Del.*, vol. 15, no. 1, pp. 222–228, Jan. 2000.

[7] A. M. Gaouda, M. M. A. Salama, M. K. Sultan, and A. Y. Chikhani, "Power quality detection and classification using wavelet-multiresolution signal decomposition," *IEEE Trans. Power Del.*, vol. 14, no. 4, pp. 1469–1476, Oct. 1999.

[8] G. G. Ray,P. Chakraborty,B. K. Panigrahi,M .K. Mallick**,**"On Optimal Feature Selection Using Modified Harmony Search for Power QualityDisturbance Classification"**,***2009 World Congress on Nature & Biologically Inspired Computing (NaBIC 2009)*,Bhubhaneshwar,India

[9] D. F. Specht, "Probabilistic neural networks," *Neural Netw.*, vol. 3, no. 1, pp. 109–118, 1990.

[10] V N Vapnik, *Statistical learning theory*, Wiley, New York, 1998.

[11] Chih-Wei Hsu and Chih-Jen Lin, A Comparison of Methods for Multiclass Support Vector Machines, IEEE Transactions On Neural Networks, Vol. 13, No. 2,pp-415-425, March 2002

[12] Guang-Bin Huang, Nan-Ying Liang, Hai-Jun Rong, P. Saratchandran, and N. Sundararajan, **"On-Line Sequential Extreme Learning Machine",** *the IASTED International Conference on Computational Intelligence (CI 2005)*, Calgary, Canada, July 4-6, 2005

[13] Nan-Ying Liang, Guang-Bin Huang*, Senior Member, IEEE*, P. Saratchandran*,* and N. Sundararajan*,* "A Fast and Accurate Online Sequential Learning Algorithm for Feed forward Networks"*,* IEEE Transactions On Neural Networks, Vol. 17, pg-1411-1423 ,No. 6, November 2006.

[14] S. Mishra*,* C. N. Bhende, and B. K. Panigrahi, "Detection and Classification of Power Quality Disturbances Using S-Transform and Probabilistic Neural Network", *,IEEE Transactions on Power Delivery*, Vol. 23, Page-281-287,No. 1, January 2008

[15] I. W. C. Lee and P. K. Dash, "S-transform-based intelligent system for classification of power quality disturbance signals," *IEEE Trans. Ind.Electron.*, vol. 50, no. 4, pp. 800–805, Aug. 2003.

[16] M. V. Chilukuri and P. K. Dash, "Multiresolution S-transform-based fuzzy recognition system for power quality events," *IEEE Trans. Power Del.*, vol. 19, no. 1, pp. 323–330, Jan. 2004.

[17] M. Biswal and P. K. Dash, "Detection and characterization of multiple power quality disturbances with a fast S-transform and decision tree based classifier," *J. Digit. Signal Process.* vol. 23, no. 4, pp. 1071–1083, Jul. 2013.

[18] MATLAB. Natick, MA: Math Works, Inc., 2000.